\documentclass[3p,times,twocolumn]{elsarticle}
 \biboptions{comma,sort&compress}
 
\usepackage{graphicx}
\usepackage{amsmath}
\usepackage{here}
\usepackage{ecrc}

\volume{00}

\firstpage{1}

\journalname{Nuclear and Particle Physics Proceedings}
\runauth{Stephan Narison}

\jid{nppp}
\jnltitlelogo{Nuclear and Particle Physics Proceedings}

\usepackage{amssymb}

\usepackage[figuresright]{rotating}

\def\beq{\begin{equation}}
\def\eeq{\end{equation}}
\def\bea{\begin{eqnarray}}
\def\eea{\end{eqnarray}}
\def\bq{\begin{quote}}
\def\eq{\end{quote}}

\def\nnb{\nonumber}
\def\ga{\left(}
\def\dr{\right)}

\def\nnb{\nonumber}
\def\la{\langle}
\def\ra{\rangle}

\def\ba{\vspace*{-0.2cm}\begin{array}}
\def\ea{\end{array}\vspace*{-0.2cm}}

\def\als{\alpha_s}

\def\g2{ \la\alpha_s G^2 \ra}
\def\g3{g^3f_{abc}\la G^aG^bG^c \ra}
\def\ggg4{\la\als^2G^4\ra}

\begin{document}

\begin{frontmatter}

\title{
%
Modern status of heavy quark sum rules in QCD\,$^*$} 
 
 \cortext[cor0]{Mini-Review talk presented at QCD20 - 35 years later, 23th International Conference in QCD (27-30/10/2020,
  Montpellier - FR). }

 \author[label1]{Stephan Narison
 \corref{cor1} 
 }
   \address[label1]{Laboratoire
Univers et Particules , CNRS-IN2P3,  
Case 070, Place Eug\`ene
Bataillon, 34095 - Montpellier Cedex 05, France\\
and\\
Institute of High-Energy Physics of Madagascar (iHEPMAD), 
University of Ankatso,
Antananarivo 101, Madagascar
}
\cortext[cor1]{ICTP-Trieste  researcher consultant for Madagascar.}
\ead{snarison@yahoo.fr}

\pagestyle{myheadings}
\markright{ }
\begin{abstract}
\noindent
We briefly report the modern status of heavy quark sum rules (HQSR) based on stability criteria by emphasizing the recent progresses for determining the QCD parameters ($\alpha_s$, $m_{c,b}$ and gluon condensates)
where their correlations have been taken into account. The results\,:  $\alpha_s(M_Z)=0.1181(16)(3)$, $\overline{m}_c(\overline{m}_c)=1286(16)\, {\rm MeV},~~
 \overline{m}_b(\overline{m}_b)=4202(7)\,{\rm MeV},~ \la\alpha_s G^2\ra = (6.49\pm 0.35)\times 10^{-2}~{\rm GeV^4}$, $\la g^3 G^3\ra= (8.2\pm 1.0)\times {\rm GeV}^2\la\alpha_s G^2\ra$ and the ones from recent light quark sum rules are summarized in Table\,\ref{tab:param}.  One can notice that the SVZ value of $\la\alpha_s G^2\ra$ has been underestimated by a factor 1.6,  $\la g^3 G^3\ra$ is much bigger than the instanton model estimate, while the four-quark condensate which mixes under renormalization is incompatible with the vacuum saturation which is phenomenologically violated by a factor $(2\sim 4)$. The uses of HQSR for molecules and tetraquarks states are commented. 

\end{abstract}
\end{frontmatter}
\section{Introduction}
QCD spectral sum rules (QSSR) \`a la SVZ\,\cite{SVZa,ZAKA} have been applied 
since 41 years\,\footnote{References to original works, reviews and books prior 2004 can be found in the books and reviews\,\cite{SNB1,SNB2,SNB3}}
to study successfully the hadron properties (masses, couplings and widths) and to extract some fundamental QCD parameters ($\alpha_s$, quark masses, quark and gluon condensates,...). 

In this mini-review, we concentrate on the determinations of the previous QCD parameters from heavy quark sum rules and shortly comment on the uses of these sum rules for extracting the masses and couplings of the molecules and tetraquark states. The present review complements the ones already presented in\,\cite{SNREV15}. 

We emphasize that the analysis of the correlations of the previous QCD parameters 
as done in\,\cite{SNparam,SNparama,SN19,SNcb1,SNcb2,SNcb3,SNH18} leads to a noticeable improvement of their determinations and gives an understanding of the apparent discrepancy between some earlier estimates. 
\vspace*{-0.5cm}
\section{The heavy quark sum rules (HQSR)}
As can be seen in the SVZ original papers\,\cite{SVZa,ZAKA} and in the books\,\cite{SNB1,SNB2},
the heavy quark sum rules are of the form of exponential / Borel / Laplace (LSR) sum rules and of their ratios\footnote{The name Borel sum rule has been introduced by SVZ due to the form of the OPE in this sum rule. The name Laplace has been introduced by\,Ref.\,\cite{SNR} who have noticed that after including the PT radiative corrections, the LSR has the properties of an inverse Laplace transform. The non-relativistic version of the LSR has been extensively discussed by\,\cite{BELLa,BERTa}.}\,:
\bea
{\cal L}^c(\tau,\mu)&\equiv&\lim_ {\begin{tabular}{c}
$Q^2,n\to\infty$ \\ $n/Q^2\equiv\tau$
\end{tabular}}
\frac{(-Q^2)^n}{(n-1)!}\frac{\partial^n \hat\Pi}{ ( \partial Q^2)^n}\nnb\\
&=&\int_{M_b^2}^{t_c}dt~e^{-t\tau}\frac{1}{\pi} \mbox{Im}\Pi_H(t,\mu)~,
\nnb\\
 {\cal R}^c_n(\tau)&=&\frac{{\cal L}^c_{n+1}} {{\cal L}^c_n},
\label{eq:lsr}
\eea
or of the $Q_0^2$-Moments sum rules (MSR) and their ratios\,:
\bea
 {\cal M}^c_n(Q_0^2,\mu)&=&\frac{1}{n!}\ga\frac{\partial}{\partial q^2}\dr^n\Pi_H(q^2,m_Q^2)\vert_{q^2=-Q_0^2}\nnb\\
 &=&\int_{16m_Q^2}^{t_c}\hspace*{-0.cm}\frac{dt}{(t+Q^2_0)^n}\frac{1}{\pi} \mbox{Im}~\Pi_H(t,\mu)~,\nnb\\
 {r}^c_{n/n+p}&=&\frac{{\cal M}^c_{n}} {{\cal M}^c_{n+p}}~~: p=1,2,\dots,
 \eea
 where $m_Q$ is the heavy quark mass, $\tau$ is the LSR variable, $Q_0^2=0, m_Q^2,..$ is a free chosen scale, $n$ is the degree of moments, $t_c$ is  the threshold of the ``QCD continuum" which parametrizes, from the discontinuity of the Feynman diagrams, the spectral function  ${\rm Im}\,\Pi_H(t,m_Q^2,\mu^2)$.   $\Pi_H(t,m_Q^2,\mu^2)$ is the generic two-point correlator defined as :
 \beq
\hspace*{-0.6cm} \Pi_H(q^2)=i\hspace*{-0.1cm}\int \hspace*{-0.15cm}d^4x ~e^{-iqx}\la 0\vert {\cal T} {\cal O}_H(x)\ga {\cal O}_H(0)\dr^\dagger \vert 0\ra~.
 \label{eq:2-point}
 \eeq
$ {\cal O}_H(x)$ is the interpolating quark bilinear local current $\bar \psi_1\Gamma_{12}\psi2$ for ordinary hadrons and four-quark $(\bar \psi_1\Gamma_{12}\psi_2)(\bar \psi_3\Gamma_{34}\psi_4)$ or diquark anti-diquark $(\bar \psi_1\Gamma_{12}\bar\psi_2)(\psi_3\Gamma\psi_4)$ local current for molecules or tetraquark states. $\Gamma_{ij}$ is any Dirac matrices which specify the quantum numbers of the corresponding hadronic state (and its radial excitations) which couples to the current through its decay constant:
\beq
\la 0\vert  {\cal O}_H(x)\vert H\ra= f_H M_H^{d}~,
\eeq
where $d$ depends on the dimension of the current.  
\section{The SVZ - Expansion and beyond}
According to SVZ, the RHS of the two-point function can be evaluated in QCD within the Operator Product Expansion (OPE) provided that $ \Lambda^2\ll Q^2\equiv -q^2, m_Q^2$. In this way, it reads\,:
\beq
\Pi_H(q^2,m_Q^2,\mu)=\sum_{D=0,2,..}\hspace*{-0.25cm}C_{D}(q^2,m_Q^2,\mu)\la O_{D}(\mu)\ra~, 
\eeq
where $C_{D}$ are calculable Wilson coefficients and $\la O_{D}(\mu)\ra$ are non-perturbative gauge invariant condensates of dimension $D$. The usual perturbative (PT) contribution corresponds to $D=0$ while the quadratic quark mass corrections enter via $D=1$. 

The asymptotic behaviour of the PT series is often expected to have an exponential behaviour (Borel sum) according to the large $\beta$ approximation and then alternate signs are expected to be seen at large orders of PT. However, the known calculated terms up to order $\alpha_s^5$ of the vector correlator $D$-function do not yet show such properties. In Refs.\,\cite{CNZa,CNZb,ZAKa,ZAKb}, a phenomenological parametrization of these higher order terms  due to UV-renormalons have been proposed which is quantified in terms of a tachyonic gluon mass squared.  Its phenomenological value from $e^+e^-\to$ hadrons data\,\cite{SND21}, $\pi$-Laplace sum rule\,\cite{CNZa} and from an analysis of the lattice data of the pseudoscalar $\oplus$ scalar two-point correlators\,\cite{CNZb}\, lead to the average\,\cite{SZ}:
\beq
(\alpha_s/\pi)\lambda^2\simeq -(7\pm 3)\times 10^{-2}\,{\rm GeV}^2.
\eeq
The existence of this $D=2$ term not present in the standard OPE (absence of gauge invariant $D=2$ term) has raised some vigourous (unjustifed and emotional) reactions though its contribution is tiny in the sum rule and $\tau$-decays\,\cite{SNTAU} analyzes and that it has solved some paradoxical sum rule scale puzzles\,\cite{CNZb}. This $D=2$ term is also seen in some AdS/QCD  models\,\cite{ADS1,ADS2,ADS3}. However, this term is not of InfraRed origin like some other non-perturbative condensates but it is dual to the sum of higher order Ultra-Violet terms of the PT series as shown in\,\cite{SZ} : {\it better the series is known , lesser is the strength of this term which can vanish after some high order
terms of the PT series}.   A such term is dual to a geometric sum of the coefficients of the PT series and is consistent with the values of the known coefficients. 

Up to dimension-six ($D=6$), one has successively the $\la\bar \psi\psi\ra$ quark, $ \la
\alpha_sG^2\ra
\equiv \la \alpha_s G^a_{\mu\nu}G_a^{\mu\nu}\ra$ 
and $ \la g^3G^3\ra
\equiv \la g^3f_{abc}G^a_{\mu\nu}G^b_{\nu\rho}G^c_{\rho\mu}\ra$ gluons, 
 $g\la\bar \psi G\psi\ra
\equiv {\la\bar \psi g\sigma^{\mu\nu} (\lambda_a/2) G^a_{\mu\nu}\psi\ra}$ mixed quark-gluon
and the 
$g^2\la\bar \psi\psi\bar \psi\psi\ra$  four-quark condensates.  

The quantities
$m\la\bar\psi\psi\ra$ and the trace of the energy-momentum transfer\,: $\theta^\mu_\mu\equiv m\gamma \la\bar\psi\psi\ra +(1/4)\beta \la  G^a_{\mu\nu}G_a^{\mu\nu}\ra$ are known to be $\mu$-independent where $\gamma,\, \beta$ are the quark mass anomalous dimension and Callan-Symanzik $\beta$-function.

The renormalization of higher dimension  condensates have been studied in\,\cite{SNTARRACH} where it has been shown that $g^3f_{abc}\la G^aG^bG^c \ra$ does not mix under renormalization and behaves as $(\alpha_s)^{23/(6\,\beta_1)}$, where $\beta_1=-(1/2)(11-2n_f/3)$ is the first coefficient of the $\beta$-function and $n_f$ is number of quark flavours. 

The quark-gluon mixed condensate usually parametrized as $g\la\bar \psi G\psi\ra=M_0^2\la \bar \psi\psi\ra$  mixes under renormalization and runs as $(\alpha_s)^{1/(6\beta_1)}$ in the chiral limit $m=0$. The scale $M_0^2$ has been estimated from light baryons\,\cite{IOFFE,IOFFEa,IOFFEb,DOSCH,JAMI2,JAMI3,PIVOm} and heavy-light mesons\,\cite{SNhl}:
\beq
M_0^2=0.8(2)~{\rm GeV}^2.
\eeq

The four-quark condensate mixes under renormalization with some other ones which is not compatible with the vacuum saturation assumption used by SVZ.  Its phenomenological estimate from $\tau$-decays\,\cite{SNTAU}, $e^+e^-\to$ hadrons data\,\cite{LNT}, finite energy \,\cite{LAUNERb}  and baryon\,\cite{DOSCH,JAMI2,JAMI3} sum rules, leads to:
\beq
\rho \alpha_s\la\bar\psi\psi\ra^2\simeq 5.8(9)10^{-4}\,{\rm GeV}^6~:~~~\rho\simeq 2\sim 4~,
\eeq
 where $\rho$, indicates the deviation from factorization. 

A first step for the improvement of the estimate of the gluon condensate was the recent direct determination of 
the ratio of the dimension-six gluon condensate $\la g^3f_{abc} G^3\ra$ over the dimension-four one $\la\alpha_s G^2\ra$ using HQSR with the value\,\cite{SNcb1,SNcb2,SNcb3}:
\beq
\rho_G\equiv \la g^3f_{abc} G^3\ra/ \la \alpha_s G^2\ra=(8.2\pm 1.0)~{\rm GeV}^2,
\label{eq:rcond}
\eeq
which differs significantly from the instanton liquid model estimate\,\cite{NIKOL2,SHURYAK,IOFFE2} and may question the validity of a such model. 
Earlier lattice results in pureYang-Mills found:  $\rho_G\approx 1.2$ GeV$^2$\,\cite{GIACOa,GIACOc,GIACOd} such that it is important to have new lattice results for  this quantity. Note however, that the value given in Eq.\,\ref{eq:rcond} might also be an effective value of all unknown high-dimension condensates not taken into account in the analysis of \,\cite{SNcb1,SNcb2,SNcb3} when requiring the fit of the data by the truncated OPE if, at that order, the OPE does not converge. Moreover, we shall see here and in different examples that the effect of the $ \la g^3f_{abc} G^3\ra$ term is a small correction at the stability  region where the optimal results are extracted.

\section{Spectral function}
In the absence of complete data, the minimal duality ansatz\,:
\beq
\hspace*{-0.75cm}\frac{1}{\pi}\mbox{ Im}\Pi_H(t)\simeq  f^2_HM_H^{2d}\delta(t-M^2_H)
  \ + \
  ``\mbox{QCD cont.}" \theta (t-t_c)~
\label{eq:duality}
\eeq
is often used to parametrize the spectral function. Its accuracy has been tested in various light and heavy quark channels $e^+e^-\to \rho,J/\psi, \Upsilon,\dots$ where complete data are available\,\cite{SNB1,SNB2} and in the $\pi$-pseudoscalar channel where an improved parametrization of the $3\pi$ channel within chiral perturbation theory has been used\,\cite{BIJNENS}. Within a such parametrization, the ratio of sum rules is used to extract the mass of the lowest ground state as it is equal to its square. However, this analysis cannot be done blindly without studying / checking the absolute moments which can violate positivity for some values of the sum rule variables ($\tau, Q_0^2$)  though their ratio can lead to a positive number identified with the hadron mass squared. 
\section{Optimization criteria}

As the LSR sum rule variable $\tau$ and the degree $n$ of MSR are free parameters, one has to define some robust criteria for extracting the QCD parameters or/and resonances masses and couplings from the sum rules. 

Originally, SVZ have looked for a {\it sum rule window} where they requires more than 50\% contribution of the resonance and less than 50\% of the QCD continuum one which most of the sum rules practitioners continue to use. Later on using  the example of the harmonic oscillator and the $J/\psi$ channel, Bell-Bertlmann\,\cite{BELLa,BERTa} have shown that, the optimal result for an approximate series, is obtained at the minimum or inflexion point of the ground state mass with respect to its $\tau$-variation\,(see various examples in e.g\,\cite{SNB1,SNB2}). This criterion is more rigorous and improves the SVZ one as at the minimum or inflexion point, the SVZ requirement is automatically satisfied.

Later on, this {\it criterion of $\tau$-minimum sensitivity} has been extended to the one of the number $n$ of moments\,\cite{SNcb1,SNcb2,SNH18}, continuum threshold $t_c$\,\cite{SNB1,SNB2} and PT subtraction scale $\mu$\,\cite{SNL14,SNHL15,SNBc1,SNBc2,MOLE16,Zc} where they are considered as external unphysical variables. We shall discuss some examples below. 

\section{Initial QCD input parameters}
In the first iteration, the following QCD input parameters (mass in units of MeV) have been used\,:
\bea
&&\hspace*{-1.25cm}\alpha_s(M_\tau)=0.325^{+0.008}_{-0.016}~ ,~\la\alpha_s G^2\ra\,\,= (0.07\pm 0.04) ~{\rm GeV}^4.\nnb\\
&&\hspace*{-1.25cm}\overline{m}_c(\overline{m}_c)= (1261\pm 17)~,~
\overline{m}_b(\overline{m}_b)= (4177\pm 11)
~,
\label{eq:param}
\label{eq:mcmom}
\eea
The central value of $\alpha_s$ comes from $\tau$-decay\,\cite{SNTAU,PICHTAU}. The range covers the one allowed by PDG\,\cite{PDG,BETHKE} (lowest value) and the one from our determination from $\tau$-decay (highest value)\,\cite{SNTAU}. The values of $\overline{m}_{c,b}(\overline{m}_{c,b})$ are the average  from our recent determinations from charmonium and bottomium sum rules \,\cite{SNcb1,SNcb2}.
The value of  $\la\alpha_s G^2\ra$  almost covers the range from different determinations mentioned in Table 1 of Ref.\,\cite{SNparam}.

{\scriptsize
\begin{table}[hbt]
 \caption{Values of $\overline{m}_c(\overline{m}_c)$ and $\overline{m}_b(\overline{m}_b)$ in units of MeV coming from our most recent QSSR analysis based on stability criteria. Some other determinations can be found in PDG\,\cite{PDG}. }  
\setlength{\tabcolsep}{0.1pc}
    {\small
  \begin{tabular}{llll}
&\\
\hline
\hline
Masses&Values& Sources & Ref.    \\
\hline
$\overline{m}_c(\overline{m}_c)$&$1256(30)$  &${J/\psi}$ family&Ratios of LSR\, \cite{SNparam}\\
&$1266(16)$ &$M_{\chi_{0c}-M_{\eta_c}}$&Ratios of LSR\, \cite{SNparam}\\
&1264(6) &${J/\psi}$ family& MOM \& Ratios of MOM\, \cite{SNH18} \\
&1286(66)&$M_D$ & Ratios of LSR\,\cite{SNFB12}\\
&1286(16)&$M_{B_c}$ & Ratios of LSR \cite{SNBc2} \\
&{\it 1266(6)}&{\it Average}& \cite{SNBc2} \\
\\
$\overline{m}_b(\overline{m}_b)$&$4192(17)$&${\Upsilon}$ family&Ratios of LSR\,\cite{SNparam}\\
&4188(8) &$\Upsilon$ family &MOM \& Ratios of MOM\, \cite{SNH18} \\
&4236(69) &$M_B$&Ratios of MOM \& of LSR\,\cite{SNFB12}\\
&4213(59)&$M_B$& Ratio of HQET-LSR\,\cite{SNHQET13}\\
&4202(7)&$M_{B_c}$ & Ratios of LSR\,\cite{SNBc2}\\
&{\it 4196(8)}&{\it  Average}& \cite{SNBc2}\\
\hline\hline
\end{tabular}
}
\label{tab:hmass}
\end{table}
} 

\section{$\overline{m}_c$ and $\overline{m}_b$ from HQSR}
$J/\psi$ and $\Upsilon$ systems have been used since the original SVZ papers for extracting the 
charm and bottom quark masses. However, in these pioneer works\,\cite{SNB1,SNB2}, the definition of the mass extracted from  the analysis was ambiguous which has become clear after the use of the $\overline{MS}$-scheme running mass $\overline{m}_{c,b}(\overline{m}_{c,b})$\,\cite{SNmsbar}. The most recent update of these determinations are summarized in Table\,2 from\,\cite{SNBc2} which we reproduce in Table\,\ref{tab:hmass}. These results are compiled in the PDG data\,\cite{PDG}. 
\section{Correlation between $\overline{m}_c$ and $\overline{m}_b$ from $M_{B_c}$}

\vspace*{-0.5cm}
\begin{figure}[hbt]
\begin{center}
\includegraphics[width=7.5cm]{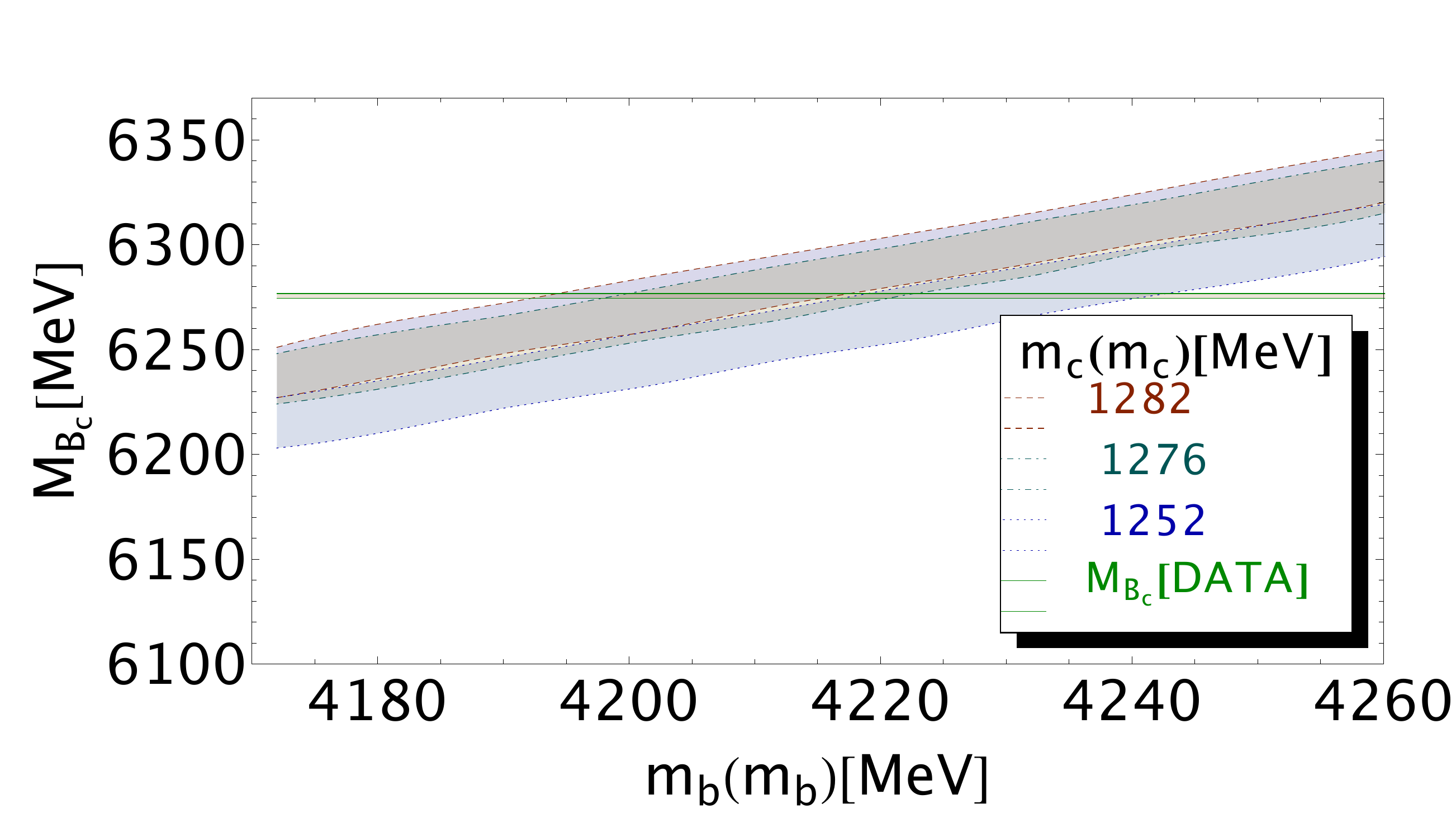}
\vspace*{-0.5cm}
 \caption{\footnotesize $M_{B_c}$ as function of $\overline m_b(\overline m_b)$ for different values of $\overline m_c(\overline m_c)$,  at the stability point  $\mu$=7.5 GeV and for the range  of $\tau$-stability values $\tau=(0.30-0.32)$ GeV$^{-2}$. }
\label{fig:mc-mb}
\end{center}
\vspace*{-0.5cm}
\end{figure} 
\section{Correlation between $\overline{m}_{c,b}$ and $\la \alpha_s G^2\ra$}
\begin{figure}[hbt]
\begin{center}
\includegraphics[width=7.5cm]{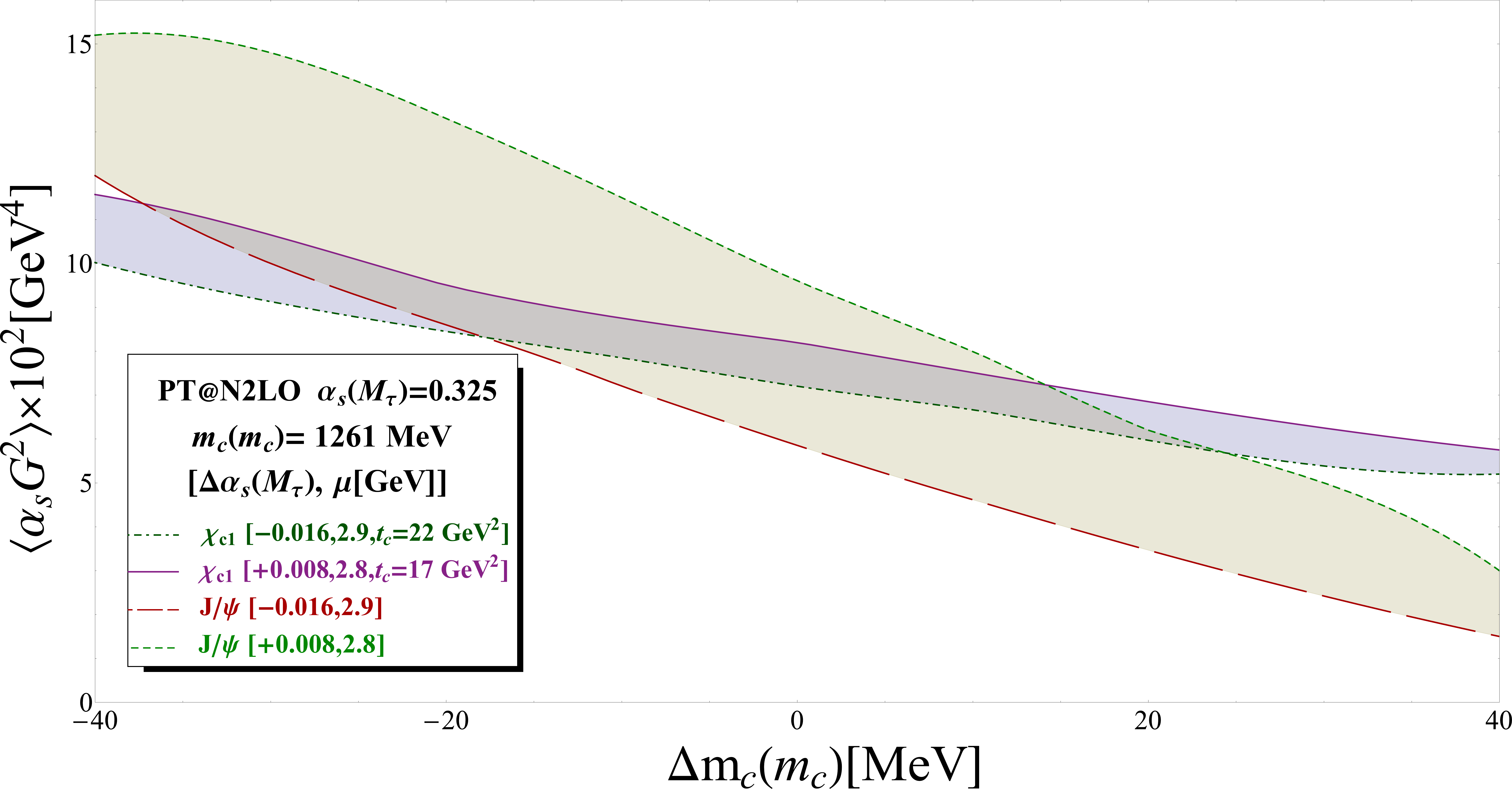}
\vspace*{-0.5cm}
\caption{\footnotesize  Correlation between $\la\alpha_s G^2\ra$ and $\overline{m}_c(\overline{m}_c)$ for given values of $\alpha_s$ and $\mu$.} \label{fig:mc-g2}
\end{center}
\vspace*{-0.5cm}
\end{figure} 

We show in Fig.\,\ref{fig:mc-mb} the correlation between $\overline{m}_c$ and $\overline{m}_b$ from the LSR analysis of $M_{B_c}$\,\cite{SNBc2}.  One can check that the numbers quoted in Table\,\ref{tab:hmass} satisfy this constraint. 

The correlation among $\overline{m}_c,\overline{m}_b$ and the gluon condensate $\la \alpha_s G^2\ra$ has been studied in details in\,\cite{SNparam,SNparama} using the Laplace sum rule (LSR) in different ordinary charmonium and bottomium channels. 
The advantage of the vector channels $J/\psi$ and $\Upsilon$ channels are that one has complete data from $e^+e^-\to J/\psi, \Upsilon,...$. For the other axial-vector (pseudo)scalar 
states, the simple duality ansatz : ``one resonance" +QCD ``continuum" gives a good description of the spectral function as tested from these $J/\psi, \Upsilon,...$ complete data (see previous quoted papers and  books). 

As an example we show in Fig.\,\ref{fig:mc-g2} the correlation between $\overline{m}_c$ and the gluon condensate $\la\alpha_s G^2\ra$. We have used the initial inputs in Eq.\,\ref{eq:param} and the value\,:
\beq
\mu_c=(2.85\pm 0.05)~{\rm GeV}, 
\eeq
of the subtraction point $\mu_c$ at the stability region for the charmonium $J/\psi$ and $\chi_{c1}$ channels from LSR\,\cite{SNparam,SNparama}.
This figure clearly shows that with the alone $J/\psi$ channel, one cannot fix accurately the value of the gluon condensate as it is sensitive to the input values of $\overline{m}_c$, where a fat band is observed.  This feature explains the apparent discrepancy between different results in the literature where only the value from the $J/\psi$ channel has been used. Adding the  $\chi_{c1}$-channel in the analysis with a narrow band improves the determination of  $\la\alpha_s G^2\ra$ (see Fig.\,\ref{fig:mc-g2}), which leads at the intersection region to\,:
\bea
\la\alpha_s G^2\ra &=& (8.5\pm 3.0)\times 10^{-2}~{\rm GeV}^4, \nnb\\
\overline{m}_c(\overline{m}_c)&=&(1256\pm 30)~{\rm MeV}~.
\eea
Using the average value of $\overline{m}_c(\overline{m}_c)$ in Table\,\ref{tab:hmass}, one can deduce from Fig.\,\ref{fig:mc-g2} the improved estimate\,:
\beq
\la\alpha_s G^2\ra = (7.35\pm 0.65)\times 10^{-2}~{\rm GeV}^4. 
\label{eq:g2-psi}
\eeq
As discussed in Ref.\,\cite{SNparam}, the inclusion of an estimated N3L0 PT and  NLO gluon condensates corrections give negligible contributions. This result can be compared with the average of different sum rules determinations prior 2017\,(see Table 1 of \cite{SNparam}):
\beq
\la\alpha_s G^2\ra = (6.25\pm 0.45)\times 10^{-2}~{\rm GeV}^4.
\label{eq:g2-mean}
\eeq
This result confrms the conclusion of\,\cite{BELLa,BERTa,LAUNERb} stating that the SVZ value 0.04 GeV$^4$ has been underestimated by about a factor 2. 
\section{Correlation between $\overline{m}_b$ and $\alpha_s$}

Some other correlations of $\overline{m}_{c,b}$ with $\la \alpha_s G^2\ra$ have been also studied by Ref.\,\cite{SNparam} in different channels but have given weaker constraints than the one presented above. Instead, these channels have provided improved estimates of $m_{c,b}$ with the results in Table\,\ref{tab:hmass}.
\begin{figure}[hbt]
\begin{center}
\includegraphics[width=7.5cm]{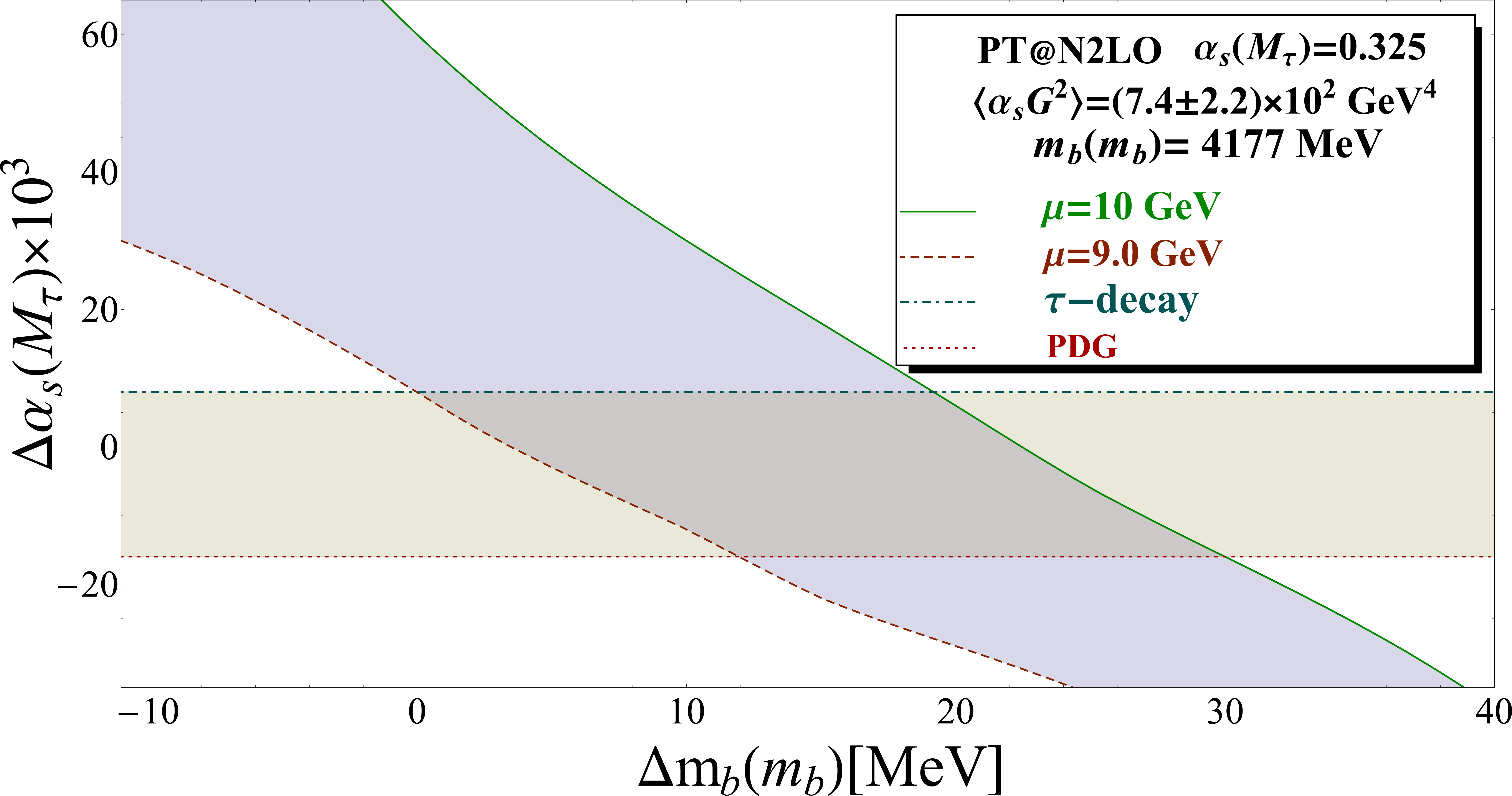}
\caption{\footnotesize  Behaviour of $\Delta\alpha_s(M_\tau)$ versus $\overline{m}_{b}(\overline{m}_{b})$ from the ratio of LSR moments ${\cal R}_{\Upsilon}$ The horizontal band corresponds to the range of $\alpha_s$ value given in Eq.\,\ref{eq:param}.}
\label{fig:mb-alfas}
\end{center}
\vspace*{-0.5cm}
\end{figure} 
In the bottomium channel, $\overline{m}_b$ is instead correlated to $\alpha_s$ as $\la\alpha_s G^2\ra$ gives a relatively smaller correction than in the case of charmonium. We show this correlation in Fig.\,\ref{fig:mb-alfas} where on can deduce that, for given values of $\alpha_s$, one can constrain the one of $\overline{m}_b$. The value:
\beq
\mu_b=(9.5\pm 0.5)~{\rm GeV}~,
\eeq
at the $\mu$-stability region has been used.
\begin{figure}[hbt]
\begin{center}
\includegraphics[width=8cm]{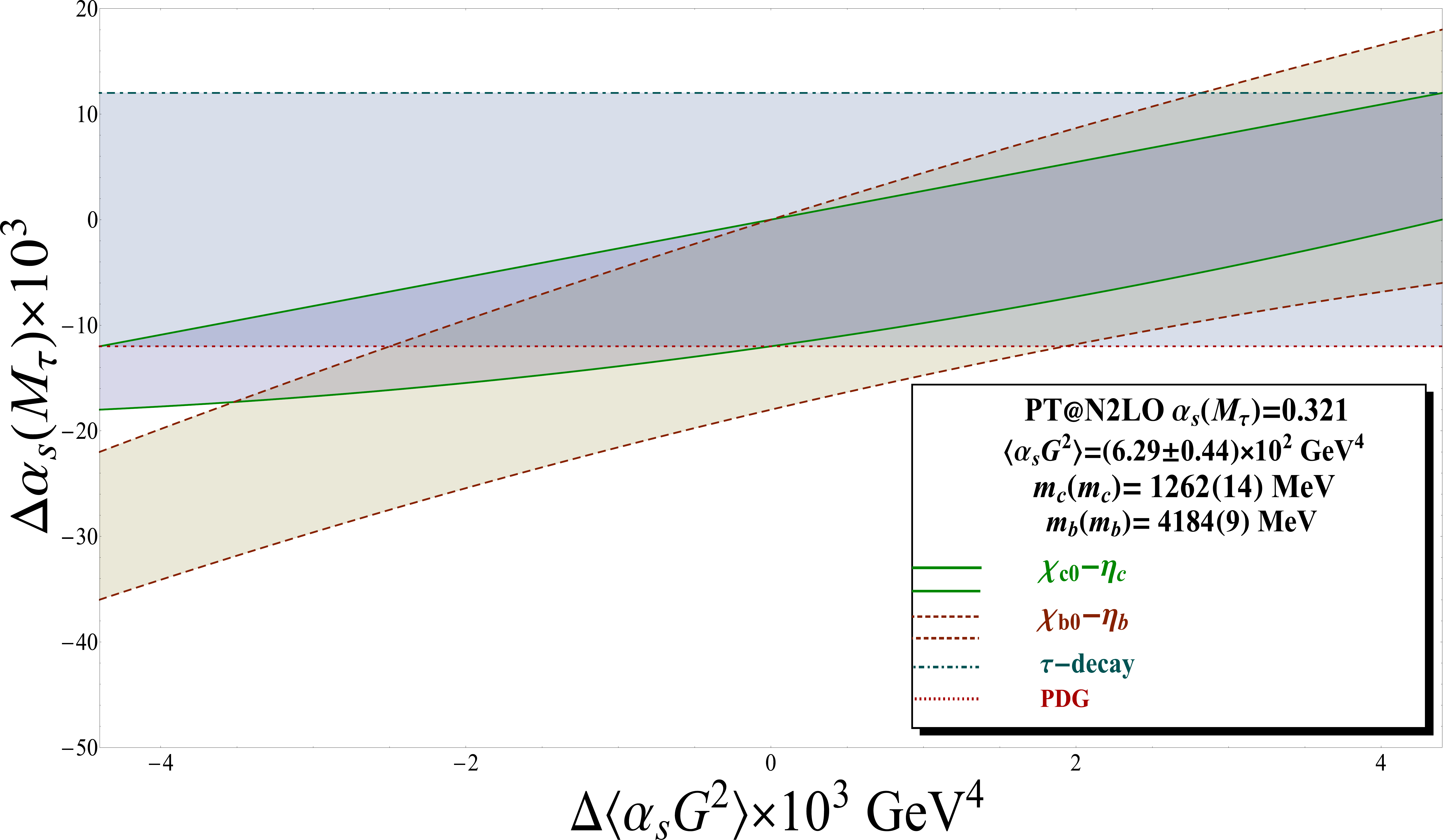}
\vspace*{-0.25cm}
\caption{\footnotesize  Correlation between $\alpha_s$ and $\la \alpha_s G^2\ra$ by requiring that the sum rules reproduce the (pseudo)scalar mass-splittings. The central value of $\la \alpha_sG^2\ra$ in the legend is the one of the previous average obtained in\,\cite{SNparam} while the masses come from Eq.\,\ref{eq:param}.} 
\label{fig:alfas-g2}
\end{center}
\end{figure} 
\begin{figure}[hbt]
\begin{center}
\includegraphics[width=8.cm]{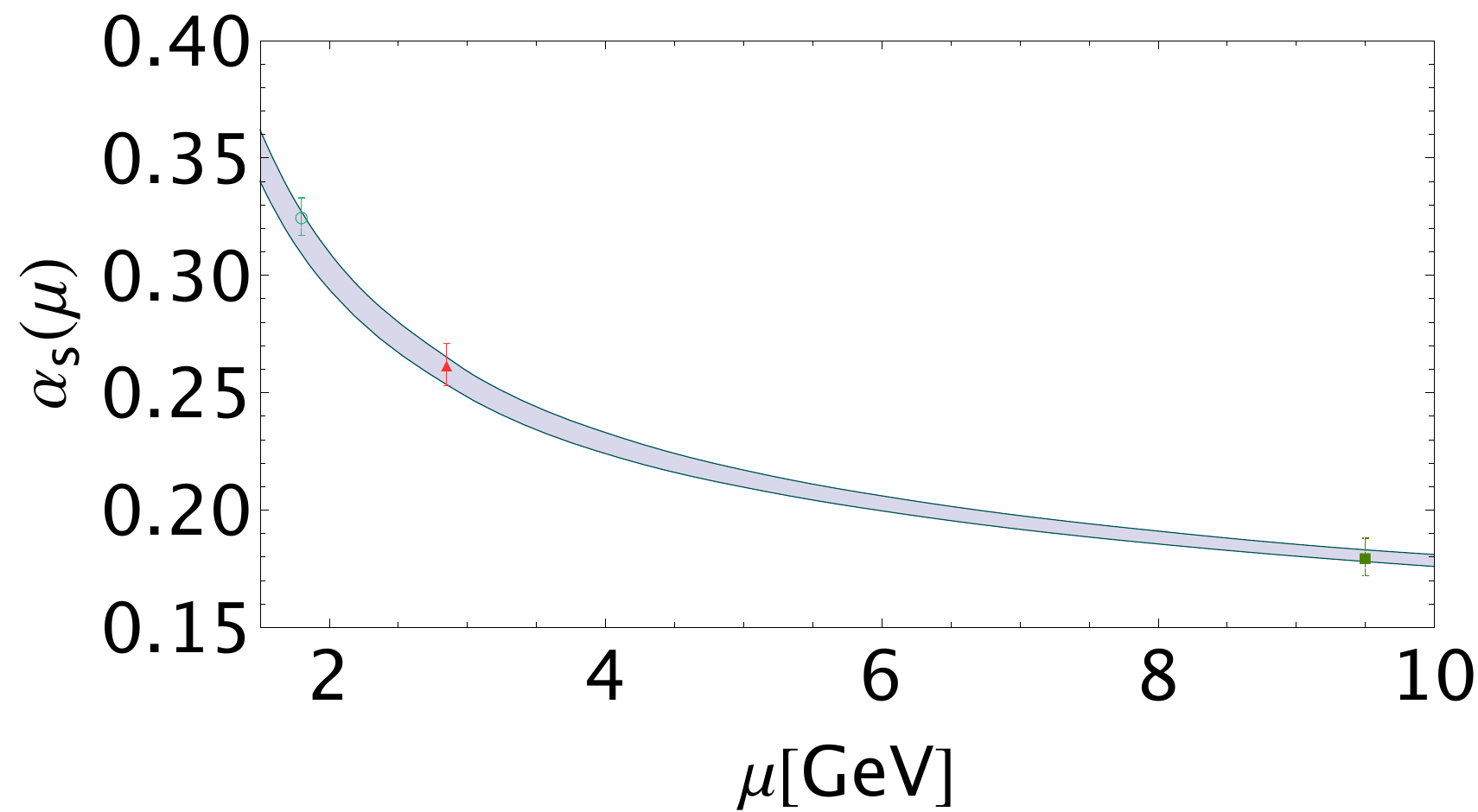}
\vspace*{-0.25cm}
\caption{\footnotesize  Comparison with the running of the world average $\alpha_s(M_Z)=0.1181(11)$\,\cite{BETHKE,PDG} of our predictions at three different scales: $M_\tau$  for the original low moment $\tau$-decay width\,\cite{SNTAU} (open circle),  2.85 GeV for $M_{\chi_{c0}}-M_{\eta_c}$ (full triangle) and  9.5 GeV for $M_{\chi_{b0}}-M_{\eta_b}$ (full square)\,\cite{SNparam}.}
\label{fig:alfas}
\end{center}
\vspace*{-0.75cm}
\end{figure} 

\vspace*{-0.5cm}
\section{$\alpha_s$ and $\la \alpha_s G^2\ra$ correlation from $M_{\chi_{0c(0b)}}-M_{\eta_{c(b)}}$}
The correlation between $\alpha_s$ and $\la \alpha_s G^2\ra$ is more pronounced from the analysis of the  $M_{\chi_{0c(0b)}}-M_{\eta_{c(b)}}$ mass-splittings.  In doing a such analysis, first, it has been checked that the LSR reproduces accurately the absolute masses of the (pseudo)scalar states $\eta_{c(b)}$ and $\chi_{0c(0b)}$\,\cite{SNparam,SNparama}. Second, the mass-splittings have been extracted directly from the sum rules while the Double Ratio of LSR\,\cite{SNB1,SNB2,DRSR} has not been used because each individual mass sum rules do not optimize at the same value of $\tau$.
The effect due to $\overline{m}_{c,b}$ and to $\mu$ in the stability regions induce respectively an error of about $(1\sim 2)$ MeV and 8 MeV. The largest effects are due to the changes of $\alpha_s$ and $\la \alpha_s G^2\ra$. We show their correlations in Fig\,\ref{fig:alfas-g2}. We have runned the value of $\alpha_s$ from $\mu_c=2.85$ GeV to $M_\tau$ in the charm channel and from $\mu_b=9.5$ GeV to $M_\tau$ in the bottom one where the values of $\mu$ correspond to the scales at which the sum rules have been evaluated:
\bea
&&\hspace*{-1cm} \alpha_s(2.85)=0.262(9) \leadsto\alpha_s(M_\tau)=0.318(15)~\nnb\\
&&\hspace*{0.25cm} \leadsto\alpha_s(M_Z)=0.1183(19)(3) ~,\nnb\\
&&\hspace*{-1cm} \alpha_s(9.50)=0.180(8) \leadsto\alpha_s(M_\tau)=0.312(27)\nnb\\
&&\hspace*{0.25cm} \leadsto\alpha_s(M_Z)=0.1175(32)(3) ~.
\label{eq:alfas1}
\eea
The last error is due to the running procedure. 
We have requested that the method reproduces within the errors the experimental mass-splittings by about ($2\sim 3$) MeV. The geometric mean of the two previous values of $\alpha_s$ is\,:
\beq
\hspace*{-0.5cm}\alpha_s(M_\tau)=0.317(15) \leadsto\alpha_s(M_Z)=0.1181(19)(3),
\label{eq:alfas}
\eeq
which is (surprisingly) in a very good agreement with the world average\,\cite{BETHKE,PDG}\,:
\beq
\alpha_s(M_Z)=0.1181(11)~.
\label{eq:asworld}
\eeq
We also show in Fig.\,\ref{fig:alfas} the running of the world average where we put the two determinations obtained at two different scales $\mu_c=2.85$ and $\mu_b=9.5$ GeV. 

Adding into the analysis the range of input $\alpha_s$ values given in Eq.\,\ref{eq:param} (light grey horizontal band in Fig.\,\ref{fig:alfas-g2}), one can deduce 
stronger constraints on the value of $\la\alpha_s G^2\ra$\,:
\beq
\la \alpha_s G^2\ra=(6.39\pm 0.35)\times 10^{-2}~{\rm GeV}^4.
\label{eq:glue2}
\eeq
Combining the previous values in Eqs.\,\ref{eq:g2-psi}, \ref{eq:g2-mean} and \ref{eq:glue2}, one obtains the {\it new sum rule average}:
\beq
\la \alpha_s G^2\ra\vert_{\rm average}=(6.49\pm 0.35)\times 10^{-2}~{\rm GeV}^4,
\label{eq:glue3}
\eeq
where we have retained the error from the most precise determination in Eq.\,\ref{eq:glue2} instead of the weighted error of 0.25. This result 
definitely rules out some eventual lower and negative values quoted in Table 1 of Ref.\,\cite{SNparam}. 
\section{QCD parameters from QSSR }
We compile in Table\,\ref{tab:param} the recent values of the QCD parameters obtained from heavy and light quarks QCD spectral sum rules within stability criteria.
We have introduced the renormalization group invariant light quark parameters $\hat m_\psi$ and $\hat \mu_\psi$\,\cite{FNR,SNB2}\,:
\bea
\overline {m}_\psi(\mu)&=&\hat m_\psi /(\log{\mu/\Lambda})^{2/-\beta_1}\ga 1+\rho_m\dr, \nnb\\
\la\overline{\bar \psi\psi}\ra(\mu)&=&-\hat \mu_\psi^3(\log{\mu/\Lambda})^{2/-\beta_1}/\ga 1+\rho_m\dr~,
\eea
with :  $\rho_m=0.8951a_s+1.3715a_s^2+0.1478a_s^3$ for $n_f=3$ light flavours\,\cite{SNB2,RUNDEC}, where $ \overline {m}_\psi$ and $\la\overline{\bar\psi\psi}\ra$ are the running mass and running condensate and $a_s\equiv \alpha_s/\pi$. 
{\scriptsize
\begin{table}[hbt]
\setlength{\tabcolsep}{0.1pc}
    {\small
  \begin{tabular}{llcc}
&\\
\hline
\hline
Parameters&Values&Sources& Ref.    \\
\hline
\it Heavy \\
$\alpha_s(M_Z)$& $0.1181(16)(3)$&$M_{\chi_{0c,b}-M_{\eta_{c,b}}}$&
\cite{SNparam,SNparama} \\
$\overline{m}_c(\overline{m}_c)$ [MeV]&$1266(6)$ &$D, B_c \oplus$&(see Table\,\ref{tab:hmass}) \\
&&$ {J/\psi}, \chi_{c1},\eta_{c}$ &\\
$\overline{m}_b(\overline{m}_b)$ [MeV]&$4202(8)$ &$B,B_c\oplus{\Upsilon}$&  (see Table\,\ref{tab:hmass}) \\
$\la\alpha_s G^2\ra$ [GeV$^4$]& $6.49(35) 10^{-2}$&Light-Heavy &
 \cite{SNparam}, this review\\
${\la g^3  G^3\ra}/{\la\alpha_s G^2\ra}$& $8.2(1.0)$[GeV$^2$]&${J/\psi}$&
\cite{SNcb1,SNcb2,SNcb3}\\
\\
\it Light \\
$\hat \mu_\psi$ [MeV]&$253(6)$ &Light &\,\cite{SNB1,SNB2,SNp15,SNLIGHT} \\ 
$\la\overline{\bar \psi\psi}\ra(2)$ [MeV]$^3$&$-(276\pm 7)^3$ &--&-- \\
$\kappa\equiv\la \bar ss\ra/\la\bar dd\ra$& $0.74(6)$&Light-Heavy&\cite{SNB1,SNB2,SNp15,SNLIGHT,HBARYON1,HBARYON2}\\
$\hat m_u$ [MeV]&$3.05\pm 0.32$&Light &\,\cite{SNB1,SNB2,SNp15,SNLIGHT} \\
$\hat m_d$ [MeV]&$6.10\pm0.57$ &-- &-- \\
$\hat m_s$ [MeV]&$114(6)$ &-- & -- \\
$\overline{ m}_u$ (2) [MeV]&$2.64\pm 0.28$ &-- &-- \\
$\overline{ m}_d$ (2) [MeV]&$5.27\pm 0.49$ &-- & -- \\
$\overline{ m}_s$ (2) [MeV]&$98.5\pm 5.5$ &-- & --\\
$M_0^2$ [GeV$^2$]&$0.8(2)$ &Light-Heavy&\,\cite{SNB1,SNB2,IOFFE,IOFFEa,IOFFEb,JAMI2,DOSCH,JAMI3,PIVOm,SNhl}\\

$\rho \alpha_s\la \bar \psi\psi\ra^2\times 10^{4}$ &$5.8(9) $[GeV$^6$] &Light,$\tau$-decay&\cite{SNTAU,LNT,LAUNERb,JAMI2,DOSCH,JAMI3}\\
\hline\hline
\end{tabular}}
 \caption{QCD parameters from  heavy and light quarks QSSR (Moments, LSR and ratios of sum rules) within stability criteria. The running light quark masses and condensates have been evaluated at 2 GeV.
 }  
\label{tab:param}
\end{table}
} 
We have not considered the values of the light quark masses and condensates when the instanton effect is included in the SVZ-expansion.  The corresponding results are also disfavoured by lattice calculations\,\cite{LATT}. 
\section{HQSR for molecules / four-quark states}
Within the last ten years, HQSR have been often used for extracting the masses and couplings of the molecules and tetraquark states from two-point functions built with quartic quark or diquark anti-diquark  currents or their widths using vertex or light cone sum rules.  Though the phenomenological results are quite successful, there are general comments on the existing works in the literaure which reminds the early days of the SVZ sum rules :

-- The analysis is often done at LO of perturbation theory where the definition of the heavy quark mass which plays a crucial role is ambiguous. The common prefered choice of different authors is the running $\overline{MS}$ running mass which is not justified at all without adding radiative corrections which are not easy to calculate. An exception is the series of works in\,\cite{MOLE16,Zc} where the factorised NLO contributions have been considered. 

-- One also notice the inflation of including higher dimension condensates (sometimes until D=12) which can be a good point. However, one should notice that only a class of diagrams have been calculated and that these high dimension condensates mix under renormalization\,\cite{SNTARRACH} which is incompatible with  the vacuum saturation often used to estimate them. The simple typical example is the four-quark condensates discussed in the introduction which mix under renormalization where a violation of the vacuum saturation by a factor $2\sim 4$ has been observed from a fit to the data. 

-- In most papers, the optimization procedure based on the sum rule window of SVZ remains handwaving as the authors have to introduced some inaccurate criteria like larger than 50\% contribution for the ground state and smaller than 50\% for the QCD continuum which, often, is not properly taken into account in the error analysis. Besides, this point the error analysis is often done in a sloppy way and not in details such that the error is difficult to be appreciated and to be checked. 

-- Many authors continue to use old estimates of the QCD parameters by SVZ which (to my opinion) SVZ themselves will not consider seriously at present . One should be aware that  a lot of efforts to improve the values of these parameters and the field during many decades have been done and should not be ignored. Indeed, reading most of the present papers, one has the impression that no progress on the improvment of the method has been done since its discovery and the clock has stopped in 1979.

\section{Conclusions}
We have shortly reviewed the modern status of heavy quark sum rules (HQSR) where we have emphasized the progresses on the determinations of the QCD parameters (quark masses, gluon condensates and QCD coupling $\alpha_s$) which have been achieved thanks to the analysis of the correlations among these parameters. 

We have also commented the present uses of HQSR for exotic hadron molecules and four-quark states. 

To my personal opinion, QCD spectral sum rules can still have a long lifetime for studying successfully the properties of hadrons and for extracting the QCD parameters, However, provided we continuously improve it by doing a more careful job ! 


\end{document}